\documentstyle[aps,prl,epsfig,twocolumn,floats]{revtex}

\def\npb{Nuc. Phys. B}
\def\prd{Phys. Rev. D}

\def\ie{{\em i.e.\ }}

\def\rmmat#1{{\hbox{\rm #1}}}
\def\rmscr#1{\rmmat{\scriptsize #1}}
\begin{document}
\draft
\tighten
\twocolumn[\hsize\textwidth\columnwidth\hsize\csname@twocolumnfalse%
\endcsname

\title{Vacuum Decay Constraints on a Cosmological Scalar Field}
\author{Jeremy S. Heyl \& Abraham Loeb} 
\address{Harvard-Smithsonian Center for Astrophysics, MS-51, 60
Garden Street, Cambridge 02138}

\date{\today}
\maketitle

\begin{abstract}
If the potential of a scalar field $\phi$ which currently provides the
``dark energy'' of the universe, has a negative minimum $-M_0^4$, then
quantum-mechanical fluctuations could nucleate a bubble of $\phi$ at a
negative value of the potential. The bubble would expand at the speed of
light. Given that no such bubble enveloped us in the past, we find that any
minimum in $V(\phi)$ must be separated from the current $\phi$ value by
more than $\min \{1.5 M_0, 0.21 M_{\rm Pl}\}$, where $M_{\rm Pl}$ is the
Planck mass.  We also show that vacuum decay renders a cyclic or ekpyrotic
universe with $M_0^4\gtrsim 10^{-10} M_{\rm Pl}^4$, untenable.

\end{abstract}

]
\narrowtext

Observations of Type Ia supernova \cite{Garnavich1998} and microwave
background anisotropies \cite{deBernardis2000} indicate that the universe
is currently dominated by ''dark energy'' with a negative pressure.  One of
the simplest realizations of such energy is in the form of a nearly
massless cosmological scalar field $\phi$ (``quintessence'') which is
rolling down a shallow potential $V(\phi)$ and leading to an accelerated
expansion of the universe at the current epoch.  {\it Can this potential
take any arbitrary form for large deviations of $\phi$ from its present
value $\langle\phi\rangle$ ?} In this {\it Letter} we place interesting
constraints on the shape of $V(\phi)$ by considering the possibility of
vacuum decay.  If $V(\phi)$ has a negative minimum, then a rare
quantum-mechanical fluctuation in $\phi$ could nucleate a bubble inside of
which the energy gained from $V(\phi)$ is larger than the energy invested
in the gradients of $\phi$ on the bubble walls. Once nucleated, the walls
of the bubble will propagate outwards similarly to a relativistic burning
front and eventually envelope the entire volume within its future light
cone, transforming $\phi$ to a lower energy state.  A universe in which
such nucleation events occur would become highly inhomogeneous and would
evolve differently than expected based on the smooth semi--classical
trajectory of $\phi$.

Bubble nucleation could have fatal consequences for some cosmological
models. For example, it has recently been proposed \cite{Steinhardt01} that
a potential which reaches a minimum within a narrow and deep ``pit'' in
Planck mass units ($M_\rmscr{Pl}=1.2\times 10^{19}~{\rm GeV}$) can drive a
cyclic universe with an infinite sequence of `big bang'--`big crunch'
cycles along its history. We will show that if the potential has a
non-negligible slope in Planck mass units, then the vacuum in such a
universe will inevitably become unstable and highly inhomogeneous during
its big crunch. More generally, we will attempt to derive a condition for
the stability of the vacuum in a cosmology with an arbitrary $V(\phi)$.  We
start by analysing the formation of a single bubble.

\paragraph*{Bubble formation.}
We consider a cosmological scalar field with a general potential $V(\phi)$
that varies smoothly (without a barrier) towards a minimum value of
$-M_0^4$ in a pit centered at $\phi=\phi_\rmscr{min}$ (see Fig. 1; we use
units of $c=\hbar=1$).  This generic shape has been suggested in the
context of a collision between two brane worlds which approach one another
along an extra dimension (see [3--5] and references therein).  The
potential is relatively flat for $\phi > \phi_\rmscr{c}$ but declines
rapidly for $\phi_\rmscr{min} < \phi < \phi_\rmscr{c}$.

\begin{figure}
\centerline{\epsfig{file=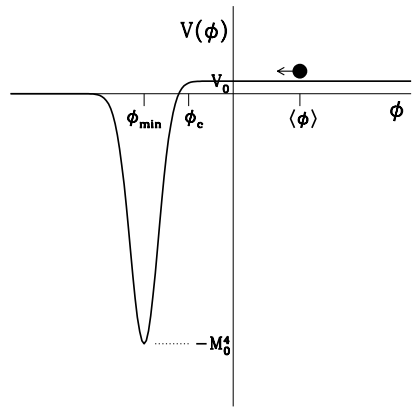, height=2.5in}}
\caption{The potential of a cosmological scalar field under consideration.
The potential obtains a minimum of $-M_0^4$ at $\phi_\rmscr{min}$ which is
much deeper than its current positive value $V_0$ at $\langle
\phi\rangle$. This shape was inspired recently by string theory (see [3-5]
and references therein). }
\end{figure}

Classically, the field is expected to reach the potential minimum
throughout the universe in the future. The current potential value of $V_0$
at $\phi=\langle \phi\rangle$ (corresponding to a cosmological density
parameter $\Omega_\Lambda=0.7$) is $\sim 10^{-123}M_\rmscr{Pl}^4$, while
$M_0$ is expected to have a non-negligible value in $M_\rmscr{Pl}$ units.
Since $V_0 \ll\ M_0^4$, we may ignore the small positive value of $V_0$ and
assume that the potential is perfectly flat for $\phi>\phi_\rmscr{c}$. In
order to treat the large derivative of $V(\phi)$ at $\phi<\phi_\rmscr{c}$
with the smallest number of free parameters, we approximate the potential
as linear in this regime and consider
\begin{equation}
V(\phi) = 
-M_0^4 \frac{\phi -
\phi_\rmscr{c}}{\phi_\rmscr{min}-\phi_\rmscr{c}} 
\Theta(\phi_\rmscr{c}-\phi) ,
\label{eq:linear}
\end{equation}
where $\Theta (x)$ is the Heaviside step function.  For reasons that will
become apparent below we are not concerned with the behavior of $V(\phi)$
at $\phi<\phi_\rmscr{min}$.

Lee and Weinberg \cite{Lee86} calculated the probability of forming a
bubble within which $\phi < \langle \phi \rangle$ for the potential in
Eq.~(\ref{eq:linear}), neglecting the effects of gravity.  We have verified
that the positive contribution of the gravitational energy can be
self--consistently neglected in their solution as long
as\footnote{Interestingly, even though the bubbles under consideration are
not characterized by a thin wall, we find that the condition for neglecting
gravity is similar to that obtained for potentials that satisfy the thin
wall approximation but have a non-negligible wall thickness
\cite{Coleman80,Lee87}.}~ $(\langle \phi \rangle - \phi_\rmscr{c}) \ll\
\frac{3}{8\sqrt{\pi}} M_\rmscr{Pl}$.  We focus our discussion on this
regime.

The field configuration that extremizes the Euclidean action for this
potential, \ie the ``bounce'', has the value of $\phi$ vary between $\phi =
\langle \phi \rangle$ and $\phi = 2 \phi_\rmscr{c} - \langle \phi \rangle$;
therefore, if $(\langle \phi \rangle - \phi_\rmscr{c}) < (\phi_\rmscr{c} -
\phi_\rmscr{min})$, the ``bounce'' configuration is restricted to $\phi >
\phi_\rmscr{min}$. Generically, $(\phi_\rmscr{c} - \phi_\rmscr{min}) \sim
M_\rmscr{Pl}$ \cite{Steinhardt01}, and so the condition that gravity be
unimportant is sufficient to ensure that the ``bounce'' does not go past
the minimum of the potential.

The production rate per unit volume of bubbles is \cite{Lee86}
\begin{eqnarray}
\lambda = \frac{4}{9} \pi^2 A & &  \left ( \langle \phi \rangle - \phi_\rmscr{c} \right )^4 \times \nonumber \\*
& & \exp \left \{ -\frac{32 \pi^2}{3} 
\frac { \left ( \langle \phi \rangle - \phi_\rmscr{c} \right )^3  
\left ( \phi_\rmscr{c} - \phi_\rmscr{min} \right )} {M_0^4} \right \},
\label{eq:rate}
\end{eqnarray}
where $A$ is a constant of order unity.  Once formed the bubbles expand at
the speed of light \cite{Coleman80,Lee86}, so the typical time
$t_\rmscr{tunnel}$ after which the fraction $p$ of the space still remains
outside of a bubble is \cite{Lee86},
\begin{equation}
t_\rmscr{tunnel} = \left [ \frac{3}{\pi \lambda} \ln \frac{1}{p} \right
]^{1/4}.
\end{equation}
If $t_\rmscr{tunnel}$ is shorter than the classical, cosmological rolling
time of the field $t_\rmscr{roll}$, then the evolution of the field will be
dominated by bubble formation.

While the field lies on the nearly flat part of the potential,
$t_\rmscr{roll} \sim (H_0)^{-1} \sim 10^{61} M_\rmscr{Pl}^{-1}$, where
$H_0$ is the current Hubble constant.  For $p=1/e$ we find that
$t_\rmscr{tunnel} > t_\rmscr{roll}$ if
\begin{equation}
\left ( \langle \phi \rangle - \phi_\rmscr{c} \right )^3 
\left (\phi_\rmscr{c} - \phi_\rmscr{min} \right ) > 5.4 M_0^4 .
\label{eq:solution}
\end{equation}
where we have crudely substituted $(\langle \phi \rangle - \phi_\rmscr{c})
\sim M_\rmscr{Pl}$~ in the pre-factor of the exponential in
Eq.~(\ref{eq:rate}).  
The constraint in Eq.~(\ref{eq:solution}) depends only logarithmically on
terms that appear outside the exponential; for example, if we take
$t_\rmscr{roll}$ to be 1 yr instead of $10^{10}$ yr then the
right-hand-side of the inequality changes its value only slightly to $4.5
M_0^4$.  As argued for standard inflation \cite{Hawking82}, if
$t_\rmscr{tunnel} < t_\rmscr{roll}$ then the resulting universe will be
highly inhomogeneous.

In conclusion, we find that the formation of a bubble inside the false
vacuum of an evolving scalar field freezes out the quantum fluctuations of
the field.  During the current epoch of cosmic acceleration, the scale of
the horizon is $\sim 61$ orders of magnitude greater than that of a
Planck--scale bubble that could begin to grow.  If the largest quantum
fluctuation within the horizon is sufficiently large, a critical bubble
will form and expand to contain its future light cone.  The largest
fluctuation within our horizon is typically a $\sim 30$--$\sigma$ event,
i.e. it is astronomically rare, but if this fluctuation is large enough to
create an expanding bubble, then the fluctuation is frozen out and renders
the universe highly inhomogeneous.

\paragraph*{Consequences today.}
Since the Earth has not been enveloped inside a bubble where the scalar
field lies on the ``cliff'' of the potential and has a negative energy
density, we find from Eq.~(\ref{eq:solution}) that today
\begin{equation} 
\left(\langle\phi \rangle - \phi_\rmscr{c}\right) > \min \left
\{ {1.75 M_0^{4/3} 
\over \left (\phi_\rmscr{c} -
\phi_\rmscr{min} \right )^{1/3}}, \frac{3}{8\sqrt{\pi}}
M_{\rm Pl} \right \} .
\end{equation}
This constrains the mean slope of the potential for $\phi > [\langle \phi
\rangle - 3/(8 \sqrt{\pi}) M_\rmscr{pl}]$ to be less than $0.002
M_\rmscr{Pl}^3$.  It also implies that $(\langle \phi\rangle -\phi_{\rm
min})> \min \{1.5 M_0, 0.21 M_{\rm Pl}\}$.

Next we demonstrate the significance of our results in the context of 
specific cosmological models.

\paragraph*{Implications for a cyclic or an ekpyrotic universe.}
Steinhardt \& Turok \cite{Steinhardt01} have recently proposed a model for
a cyclic universe based on a potential similar to that illustrated in
Fig. 1; the universe in this model is currently dominated by an effective
scalar field and is beginning to inflate. They argue that if the scalar
field passes through its global minimum after $\sim 100$ $e$-foldings, then
the universe could recollapse and bounce uniformly.  After $\sim 100$
$e$-foldings, a typical Hubble volume would be completely devoid of matter
or debris from the previous cycle; however, the quantum fluctuations in the
scalar field just before the recollapse would be able to seed the density
perturbations in the next cycle of cosmic evolution \cite{Khoury_pert}.

If we consider a simple harmonic potential for the scalar field at values
$\phi >\phi_{\rm c}$ and take the density of the flat universe today to be
$30\%$ matter and $70\%$ vacuum energy, we find through a numerical
integration that the scalar field reaches the potential minimum after $\sim
92$ $e$-folds.  Looking back in time from the moment of recollapse, the
value of $\vert\langle\phi\rangle-\phi_\rmscr{c}\vert$ at the beginning of
the final $e$-fold is $3\%$ of $M_\rmscr{Pl}$.  If the potential is too
steep a bubble will form within the past light cone and will be able to
envelope the entire Hubble volume before the scalar potential reaches its
minimum.  For the scalar field to roll rather than tunnel, the slope past
the cliff is bounded by
\begin{equation}
\frac{M_0^4}{\phi_\rmscr{c} - \phi_\rmscr{min}} < 
5 \times 10^{-6} M_\rmscr{Pl}^3 .
\end{equation}
This limit cannot be evaded without increasing the rate of change of
$\phi$, which would reduce the number of $e-$foldings before the field
reaches the minimum.  As $\langle\phi\rangle$ approaches the potential
minimum, more bubbles could form and expand if the depth of the potential
minimum is not too small.  The formation of these bubbles destroys the
homogeneity of the causally connected patch of the universe.  In the cyclic
universe the expansion will decelerate about one billion years before the
potential reaches its minimum, because at this time the kinetic energy of
the field begins to dominate the potential energy.  Based on a numerical
integration of the evolution equation for $\phi$, we find that expanding
bubbles will form at this stage unless 
\begin{equation}
\frac{M_0^4}{\phi_\rmscr{c} - \phi_\rmscr{min}} < 9 \times 10^{-11}
M_\rmscr{Pl}^3. 
\label{eq:condition}
\end{equation}
This constraint requires fine-tuning since the generic value of the slope
lies in the Planck regime.

If condition (\ref{eq:condition}) is not satisfied as the field approaches
the cliff in the potential (where $dV/d\phi$ changes dramatically),
expanding bubbles of negative energy density will begin to form and destroy
the homogeneity of the collapse.  The transition of the universe from an
expanding phase to a contracting phase would not be uniform but rather
proceed as a first-order phase transition with bubbles of contracting
(anti-de Sitter) spacetime appearing in the expanding background.  The
walls of the internally contracting bubbles expand at the speed of light.
Furthermore, because the interior of the bubble will deviate from spherical
symmetry, possibly as a result of amalgamating a smaller bubble or
inevitably from quantum fluctuations, the collapse within the bubble will
be chaotic and singular.  How these bubbles overlap and how chaotic their
subsequent collapse is, will depend on the precise details of the phase
transition.  An orderly collapse and reexpansion cannot be achieved
naturally in the context of the cyclic model. (For bubble dynamics in a
rapidly inflating universe, see \cite{Hawking82}.)  We note that in
similarity to other tunneling problems, the exponentially--suppressed
quantum-mechanical jump to a new, fully nonlinear configuration of $\phi$
may not be evident in any order of perturbation theory that was discussed
previously in the literature \cite{Khoury_pert}.

Our conclusions apply also to the ekpyrotic universe
\cite{Khoury01,Linde01} in which the big bang was preceded by a big crunch
through a potential similar to that illustrated in Fig. 1 but with
$V_0=0$. Since the small value of $V_0$ was ignored in our discussion, the
above results apply to this case as well.  In the ekpyrotic scenario, the
universe exists for a sufficiently long time to smooth out inhomogeneities
causally before collapsing in a big crunch and rebounding in a big bang.
The constraints derived here apply also to this scenario since many bubbles
of collapsing spacetime would form before the scalar field rolls down the
potential classically and initiates the big crunch globally.

We have found that cosmological scenarios which require the universe to
contract and rebound suffer from a first-order phase transition which
necessarily destroys the homogeneity of the universe.  Perhaps a
sufficiently finely tuned potential for the scalar field which drives the
evolution can avoid this first-order phase transition but it may not
possess the other attractive features of the ekpyrotic and cyclic
propositions.  More generally, we have constrained the future classical
evolution of the observed cosmological constant by appealing to the
apparent lack of bubbles of collapsing spacetime in the universe today.

\bigskip
\paragraph*{Acknowledgments.}
We thank Paul Steinhardt and Erick Weinberg for useful discussions.
J.S.H. has been supported by the Chandra Postdoctoral Fellowship Award \#
PF0-10015 issued by the Chandra X-ray Observatory Center, which is operated
by the Smithsonian Astrophysical Observatory for and on behalf of NASA
under contract NAS8-39073. This work has also been supported in part by
grants of A.L. from NASA (NAG 5-7039, 5-7768), and NSF (AST-9900877,
AST-0071019).

\end{document}